# Order-disorder Effects on Equation of State in FCC Ni-Al Alloys

H. Y. Geng,[1,2] M. H. F. Sluiter,[3] and N. X. Chen[1,4]

[1]*Department of Physics, Tsinghua University, Beijing 100084, China*

[2]*Laboratory for Shock Wave and Detonation Physics Research,*
*Southwest Institute of Fluid Physics,*
*P. O. Box 919-102, Mianyang Sichuan 621900, China*

[3]*Institute for Materials Research, Tohoku University, Sendai, 980-8577 Japan*

[4]*Institute for Applied Physics, University of Science and Technology, Beijing 100083, China*

## Abstract

Order-disorder effects on equation of state (EOS) properties of substitutional binary alloys are investigated with the cluster variation method (CVM) based on *ab initio* effective cluster interactions (ECI). Calculations are applied to the fcc based Ni-Al system. Various related quantities are shown to vary with concentration around stoichiometry with a surprising "W shape", such as the thermal expansion coefficient, the heat capacity and the Grüneisen parameter, due to configurational ordering effects. Analysis shows that this feature originates from the dominated behavior of some elements of the inverse of Hessian matrix. For the first time we point out that the strong compositional variation of these properties might be partially responsible for local fractures in alloys and mineral crystals under heating, highlighting the importance of subtle thermodynamic behavior of order-disorder systems.

## I. INTRODUCTION

The equation of state (EOS) is a primary but important property to understand materials behavior. Although the theory of the EOS for elemental substances is well-developed in both the ordinary density[1,2] and the abnormal density region,[3,4] its extension to alloys and compounds is a rather recent development[5] and some interesting results have been obtained.[6] It has been understood that ordering and disordering process have considerable effects on phase stability and thermodynamic behaviors of materials, as well as on the EOS, of course. For example, the pressure is increased considerably due to the order-disorder transition along the Hugoniot in $Ni_3Al$.[6] However, this effect on the EOS was investigated only at constant composition. Initial calculations have pointed to surprising compositional variations in the heat capacity.[7,8] Though these calculations dealt with simple models and some important contributions were ignored, a theoretical analysis showed that the so-called "W shape" of heat capacity around stoichiometric compositions is a general feature of ordered alloys[9] and similar phenomena can be expected for other thermodynamic quantities.

First-principles calculations based on density functional theory (DFT) have received much attention for the study of alloy phase stability with contributions from chemical effects[10] and lattice vibrations.[11] For the EOS, first-principles results are not as accurate as might be expected,[12] mainly because of the large error in calculating the bulk modulus of transition metals and the difficulty to accurately account for lattice vibrations and local distortions. However, the precision of current *ab initio* results is high enough for making definite predictions, and will be employed in this work to derive effective cluster interactions (ECI).

For a full understanding of the properties of alloys, knowledge of formation free energy is not completely sufficient. The knowledge of the EOS is essential for understanding mechanical and thermodynamical properties during adiabatic compression and so on. Thus all Gibbs free energy contributions must be considered.[5] Combining the cluster expansion method (CEM) and the cluster variation method (CVM) provides a natural and feasible approach to evaluate the EOS of alloys and solid solutions, in which configurational effects are included explicitly. The effects of ordering and disordering process can be modelled directly in this framework by variation principle of minimizing Gibbs function with respect to volume and correlation functions. It is necessary to point out that unlike vibrational and electronic excitations, excitations associated with short- or long ranged order have large en-ergy barriers so that non-equilibrium states are easily reached. Therefore, it is quite suitable to separate out the effect of ordering on the thermal properties.

In this paper we calculate order-disorder effects on the EOS and related thermal quantities for binary fcc Ni-Al alloys using *ab initio* chemical- and lattice vibrational contributions. The methodology of our calculations is discussed briefly in the next section. The model to approximate the contribution of lattice vibrations is described and ordering corrections to the thermal expansion coefficient, heat capacity at constant pressure, Grüneisen parameter and so forth are derived and calculated. The implications of the strong composition dependence are discussed.

## II. METHODOLOGY

For substitutional binary alloys, the Gibbs free energy can be written as

$$\hat{G} = \sum_i \left[ [v_i(V) + w_i(V,T)] \cdot \xi_i + k_B T \sum_{\alpha_i} a_{\alpha_i} \text{Tr}_{\alpha_i} \rho_{\alpha_i} \log \rho_{\alpha_i} + PV \right], \qquad (1)$$

where the summation is over all types of clusters, $a_{\alpha_i}$ is the möbius inversion coefficient of cluster $\alpha_i$ of type $i$ which satisfies $a_{\alpha_i} = \sum_{\beta \supset \alpha_i}{}' (-1)^{|\beta/\alpha_i|}$, the prime indicates the summation is restricted by maximal clusters; $\rho_{\alpha_i}$ is the density matrix of cluster $\alpha_i$ and is related to correlation functions $\xi_i$ by

$$\rho_{\alpha_i} = \rho^0_{\alpha_i} \left( 1 + \sum_{\substack{\beta_j \subset \alpha_i \\ \beta_j \neq \varnothing}} \xi_j \sigma_{\beta_j} \right), \quad \rho^0_{\alpha_i} = 2^{-|\alpha_i|}.$$

Here $\sigma_{\beta_j}$ is the cluster occupation variable and $|\alpha_i|$ is the number of sites contained in the $\alpha_i$ cluster.[13] The chemical and vibrational effective cluster interactions (ECI) $v_i(V)$ and $w_i(V,T)$ are derived by a generalized Connolly-Williams procedure[14] with cohesive energies and vibrational free energies of a set of superstructures,

$$v_i(V) = \sum_S \left(\xi_i^S\right)^{-1} E^S(V), \qquad (2)$$

$$w_i(V,T) = \sum_S \left(\xi_i^S\right)^{-1} F_v^S(V,T). \qquad (3)$$

The superscript $S$ denotes the superstructures and $\left(\xi_i^S\right)^{-1}$ is the general pseudo-inverse, i.e., the Moore-Penrose inverse of the correlation function matrix, which gives the least squares solution for overdetermined systems of equations.[15]

The vibrational free energy is described approximately by Debye-Grüneisen model

$$F_v(V,T) = 3k_B T \ln\left(1 - \exp\left(-\Theta_D/T\right)\right) - k_B T D(\Theta_D/T) + \frac{9}{8}k_B \Theta_D, \quad (4)$$

where $k_B$ is Boltzmann's constant and $D$ is the Debye function. The Debye temperature is approximated as[16]

$$\Theta_D = [c \cdot d_{Al} + (1-c) \cdot d_{Ni}] \left[\frac{BV^{\frac{1}{3}}}{M}\right]^{1/2}, \quad (5)$$

where $B$ is the bulk modulus as determined from the cohesive energy curve, $M$ is the atomic weight and $c$ is the concentration of Al. Scaling factors $d_{Al}$ and $d_{Ni}$ are determined from experimental $\Theta_D$'s of constituent elements at ambient condition, respectively (423K for Al and 427K for Ni), to remedy this model for transition metals and their alloys. It is necessary to point out that the Debye-Grüneisen model is rather crude. It does not have the capability to model the phonon density of state (DOS) at high frequencies properly, which results in inaccurate vibrational entropy difference among phases. However, this is not so serious because the contribution from the vibrational energy becomes more important than entropy for EOS calculations and the Grüneisen parameter is dependent mainly on low frequencies part of phonon DOS. A much more severe limitation of this model is that for cluster expansion Eq.(5) sometimes will become ill-defined, even breaks down completely if the atomic volume is beyond the inflection point of the cohesive energy curve where the bulk modulus $B = 0$. Ni-Al alloys exemplify this case, where at the Al-rich region, the equilibrium volume is beyond the inflection point of the Ni cohesive energy and the bulk modulus and Debye temperature cannot be defined properly. Therefore, in order to use the Debye-Grüneisen model, a certain hydrostatic pressure must be applied to reduce the size difference between Ni and Al. In this paper, a pressure of 30GPa is used.

The equilibrium Gibbs function is obtained by the variational principle

$$G = \hat{G}|_{\frac{\partial \hat{G}}{\partial \xi_i} = \frac{\partial \hat{G}}{\partial V} = 0}. \quad (6)$$

Then the EOS and other related quantities can be derived directly. In the framework of CVM+CEM, these quantities are calculated by numerical differentiation. E.g., the thermal expansion coefficient at composition $c$, temperature $T_0$, and pressure $P_0$ is evaluated using the formula

$$\alpha(c, T_0, P_0) = \frac{1}{V(c, T_0, P_0)} \left[ \frac{V(c, T_0 + \triangle T, P_0) - V(c, T_0 - \triangle T, P_0)}{2\triangle T} \right]. \quad (7)$$

Other quantities, the compressibility $\kappa$, heat capacity at constant pressure $C_P$ and isobaric EOS parameter $R$[3,4] can be calculated analogously with

$$\kappa = -\frac{1}{V} \left( \frac{\partial V}{\partial P} \right)_T, \quad (8)$$

$$C_P = \left( \frac{dH}{dT} \right)_P, \quad (9)$$

$$R = \frac{P}{C_P} \left( \frac{\partial V}{\partial T} \right)_P. \quad (10)$$

The isochoric EOS parameter (i.e., Grüneisen parameter) $\gamma$, and the coefficient of pressure $\beta$, however, must be computed indirectly via other thermal quantities because it is impossible to fix volume when the equilibrium Gibbs free energy is obtained variationally. Generally, the Grüneisen parameter can be obtained with

$$\gamma = \frac{\alpha V}{\kappa C_V}, \quad (11)$$

where the heat capacity at constant volume is given by $C_V = C_P - TPV\alpha\beta$ and the coefficient of pressure $\beta = \frac{\alpha}{P\kappa}$.

## III. CALCULATIONS AND DISCUSSIONS

### A. *Ab initio* calculations and phase diagram

Cohesive energies of some hypothetical Ni-Al fcc-based superstructures have been listed in Ref.[5]. Here the cohesive energies of some additional structures are given as computed with CASTEP[17,18] with the generalized gradient approximation (GGA)[19] for fcc lattice parameters from 2.8 to 4.8Å. The calculations are employ ultrasoft pseudopotentials[20] with a cutoff kinetic energy for planewaves of 540 eV. Integrations in reciprocal space are performed in the first Brillouin zone with a k-point grid with a maximal interval of 0.03Å$^{-1}$ as generated with the Monkhorst-Pack[21] scheme. The energy tolerance for the charge self-consistency convergence is $2\mu$eV/atom for all calculations. Cohesive energies at different lattice parameters are extracted from the total energies by subtracting the spin-polarized energies of isolated atoms. Then, they are fitted to Morse-type energy functions which are used to derive ECIs.

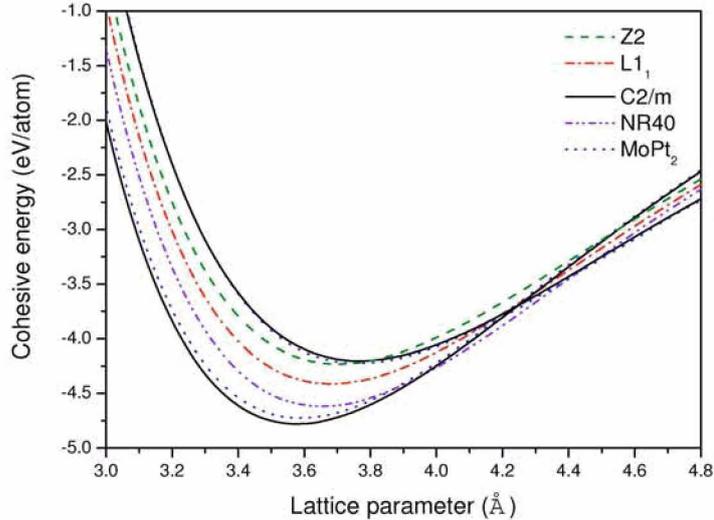

FIG. 1: Cohesive energy as function of the lattice parameter for some fcc superstructures.

TABLE I: Cohesive energies of fcc superstructures at 0 GPa.

| Structure | $c_{Al}$ | $E_{coh}$ (eV/atom) | a (Å) |
|---|---|---|---|
| C2/m | 0.333 | -4.779 | 3.578 |
| MoPt$_2$ | 0.333 | -4.725 | 3.588 |
| L1$_1$ | 0.5 | -4.412 | 3.681 |
| Z2 | 0.5 | -4.232 | 3.703 |
| NR40 | 0.5 | -4.617 | 3.652 |
| C2/m | 0.667 | -4.203 | 3.770 |
| MoPt$_2$ | 0.667 | -4.224 | 3.768 |

The calculated cohesive energy curves of the superstructures are shown in figure 1 and the equilibrium lattice parameters and cohesive energies at zero pressure are listed in table I. NR40 and C2/m are the most stable structures. We have tried to calculate the ground states and phase diagram at finite temperature with CEM and CVM approach[13,22,23] with Tetrahedron-Octahedron (T-O) approximation by inclusion of these new structures with those listed in table II of Ref.[5]. However, it fails and seems much more structures are needed. On the other hand, although T-O approximation is more accurate than T-approximation, it always cannot change the qualitative conclusions made with T-approximation. Therefore it is unnecessary to burden additional computation demands for this work when the precision of T-approximation is enough.

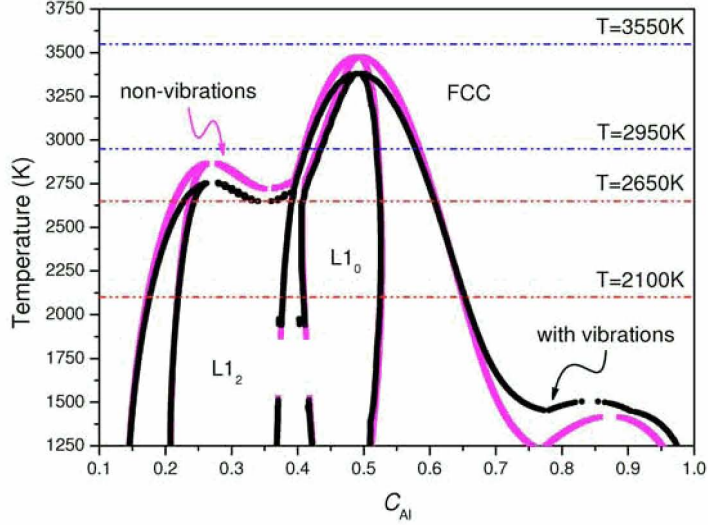

FIG. 2: The phase diagram of fcc Ni-Al in the T approximation with elastic relaxations included. For comparison, both with and without vibrational effects are shown.

After excluding the Z2 and C2/m ($Ni_4Al_2$) structures from above mentioned superstructures, a set of ECIs[24] is derived within the T-approximation that faithfully produces the correct ground states. The corresponding phase diagrams at a hydrostatic pressure of 30GPa are plotted in figure 2, where horizontal lines indicate the temperatures at which the configurational corrections of the EOS properties have been calculated as function of composition. Here global relaxation (for the elastic energy partially) is taken into account, which is responsible for the phase separation at the Al-rich side; and hydrostatic pressure is implemented to reduce the size difference between Ni and Al so that the bulk modulus of the pure phases is well defined. For comparison, the phase diagram without vibrational contribution is presented also. It is seen that order-disorder transition temperatures of $L1_2$ and $L1_0$ are slightly lowered by vibrational contributions, but less than 100K. Considering that including vibrations through anharmonity causes a volume expansion, it appears that in actuality the effect of vibrations on the order-disorder temperatures is even less. Therefore, vibrational effects on phase diagram of fcc Ni-Al appear very minor, in agreement with that inferred from first-principles calculations of the vibrational entropy.[25] However, when bcc-based structures are included, this statement might have to be reconsidered.[26]

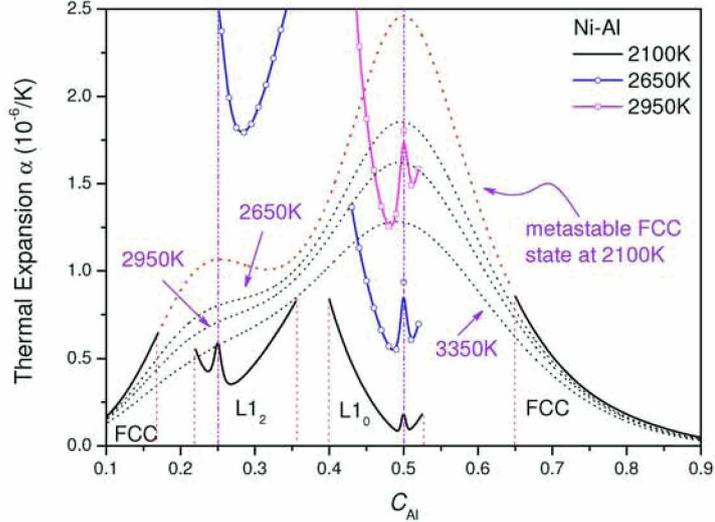

FIG. 3: Thermal expansion coefficient of Ni-Al alloys without vibrational contributions as a function of composition. Solid lines are for single phases and dotted curves are for metastable/coexisting disordered phases.

## B. Order-disorder effects on EOS quantities

Generally, thermal expansion of materials originates from anharmonic lattice vibrations. However, in the case of alloys, configurational effects are another source of thermal expansion, although its magnitude is not as large as that of vibrations. Figure 3 shows the configurational excess thermal expansion coefficient $\alpha$, computed at fixed temperatures of 2100, 2650, 2950 and 3350K, respectively, as indicated in figure 2. Stable and unstable phases and two phase regions can be found easily from figure 2. It is evident that for the metastable fcc phase, increasing temperature always decreases $\alpha$, reflecting the loss of short range order. This might suggest that ordering increases $\alpha$. However, for stable ordered phases ($L1_2$ and $L1_0$), $\alpha$ increases with temperature. This apparent contrast can be understood when realizing that disordering in the ordered state accelerates as the temperature increases. Some other details are particular interesting. The most noticeable features are the peaks and wings around stoichiometric compositions. According to Sluiter and Kawazoe[9] this is due to antisite defects near stoichiometry. The second remarkable feature is that at low temperatures, $\alpha$ of the ordered phase is much smaller than that of the fcc phase. However, when the temperature approaches the order-disorder transition temperature $T_c$, $\alpha$ in the ordered state rapidly increases and greatly exceeds the disordered $\alpha$. This leads to a sharp drop in $\alpha$ at $T_c$ when order-disorder transition is completed. Actually, according to Eq.(7) and the fact that order-disorder transitions on fcc lattice are always first order, which results in a jump of volume at $T_c$, we can conclude that $\alpha$ approaches infinite at $T_c$ and has a more steep slope on disorder side. The same conclusion is also valid for heat capacity, but not for compressibility, since pressure also jumps at $T_c$ and with Eq.(8) the compressibility has a finite value at $T_c$.

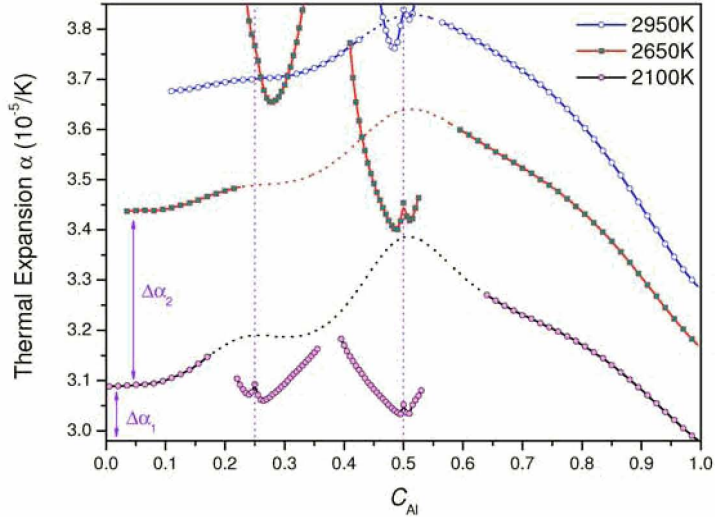

FIG. 4: Thermal expansion coefficient of Ni-Al alloys with vibrational contributions at different temperatures. The dotted lines denote the metastable/coexisting region of disordered phase.

These observations still are valid when vibrational contributions are included, as is shown in figure 4. Including vibrations now leads to $\alpha$ of the disordered phase that consistently increases with temperature. The $\alpha$ curves also become more smooth with increasing temperature. The difference between $\alpha$ of the ordered and disordered phases is enhanced a little by lattice vibrations. Simultaneously, the peaks near stoichiometric composition become less pronounced than those in figure 3. In figure 4, $\triangle\alpha_1$ is the difference between the $\alpha$ of Ni and that of Al at 2100K, and $\triangle\alpha_2$ is the increment of $\alpha$ of Ni when temperature raised from 2100K to 2650K. It is seen that their values are not so large and are comparable with the difference of $\alpha$ between ordered and disordered phases. This kind of variation of $\alpha$ (curves in FIG.4) as a function of composition and temperature due to ordering/disordering process has not been reported before. Considering that actual materials generally are not perfectly single phases with homogeneous composition, the strong composition dependence of the thermal expansion coefficient might contribute to thermal stresses in alloys (the same conclusion is also valid for mineral crystals).

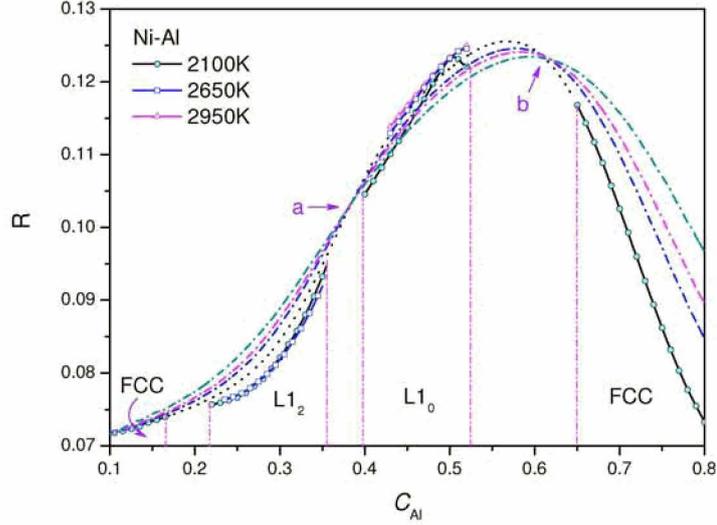

FIG. 5: Isobaric EOS parameter R as a function of Al concentration, without vibrational contributions. Dash-dotted lines indicate those of the fcc phase at 2650, 2950 and 3550K, respectively.

The EOS parameter $R$ (for isobaric) and $\gamma$ (Grüneisen parameter, for isochoric) are also modified by configurational corrections. The variation of the former with Al concentration is plotted in figure 5. The Grüneisen parameter has a shape very similar to the EOS parameter $R$ when vibrational contributions are excluded. It is evident that the effect of short-range ordering is very strong. In contrast, long-range ordering corrections are very limited, just a slightly lower (higher) $R$ in the $L1_2$ ($L1_0$) single phase region.

Remarkably, there are two points ($a$ and $b$ in FIG.5) where $R$ appears constant with temperature for the metastable fcc phase at both sides of the $L1_0$ phase. However, these points do not occur when lattice vibrations are included and we believe they have little significance for materials behavior.

When lattice vibrations are taken into account the behavior of the isochoric and isobaric

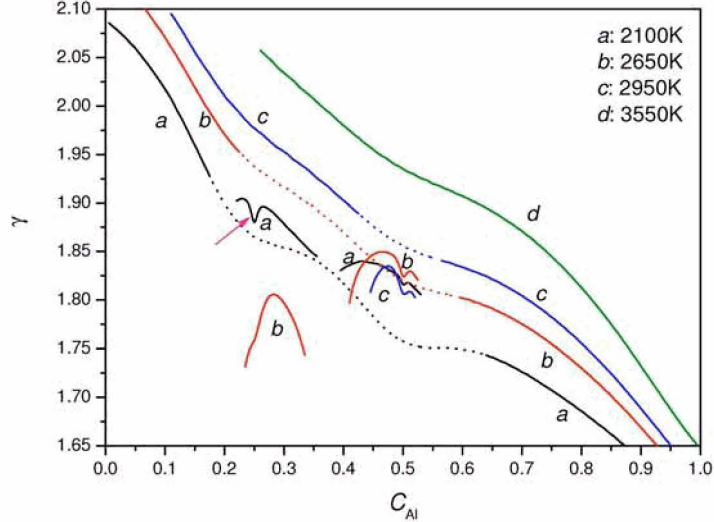

FIG. 6: The Grüneisen parameter as a function of Al concentration at different temperatures. Vibrational contributions are included.

EOS parameters changes significantly. Figure 6 shows the isochoric EOS parameter as a function of the Al concentration at different temperatures. The cross points $a$ and $b$ in figure 5 are removed by vibrational effects. The upset "W-shape" (pointed out by arrow) appears near stoichiometry. Although $\gamma$ of the $L1_0$ is rather temperature independent, $\gamma$ of the $L1_2$ phase is not. This is probably due to the order-disorder transformation of the $L1_2$ phase in the displayed temperature range. Above the order-disorder temperature $\gamma$ attains a higher value again, as the 2950K data shows (line $c$ in figure 6). This kind of rapid change of the Grüneisen parameter explains the sudden increase in pressure during an order-disorder transition in $Ni_3Al$.[6]

The heat capacity at constant pressure $C_P$ has a very similar shape as that of the thermal expansion coefficient. It suggests there is a common underlying cause. The variation of $C_P$ with Al concentration and temperature is shown in figure 7.[27] Its variation with concentration (including the peaks and wings) and with temperature is considerable, which would enhance the inhomogenous temperature distribution during thermal treatment of alloys.

The compressibility $\kappa$ is an important property to model compression behavior of materials under high pressures. It is related to the bulk sound velocity via a thermodynamic relation. Order/disorder has little effect on $\kappa$. It is slightly lower in the ordered phases than in the disordered fcc phase, as shown in figure 8. Thus, long-range order has little influence on the bulk sound velocity. However, a deviation from linearity due to short-range order is apparent. For bcc lattice, it must be pointed out that the thing is somewhat different. Figure 9 shows the square of bulk sound velocity for B2 and disordered bcc phases, where the stability of B2 phase is relative to bcc structure and has an overestimated order-disorder temperature. One can see there both strong short- and long-range order effects are presented, whereas the "W-shape" is still absent.

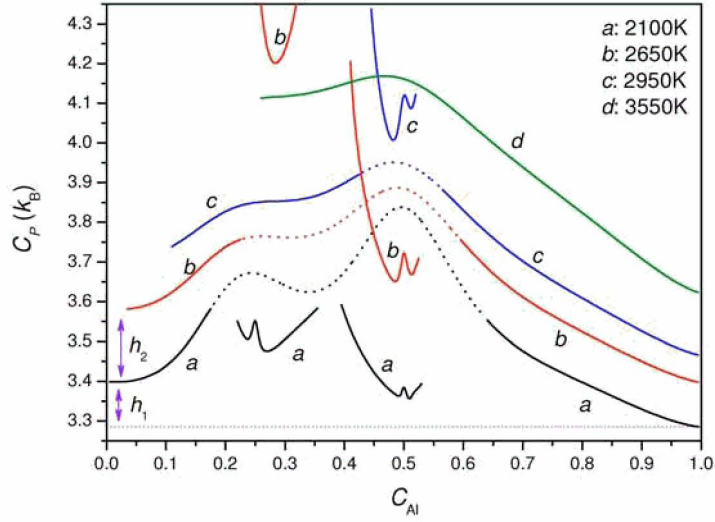

FIG. 7: The heat capacity at a constant pressure of 30GPa for Ni-Al alloys with vibrational contributions included. Notice the similarity with the thermal expansion coefficient.

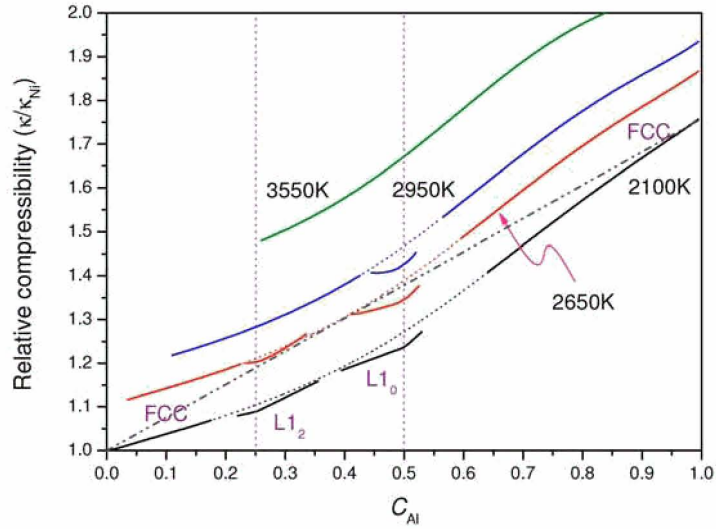

FIG. 8: The compressibility at 30GPa of Ni-Al alloys with vibrational contributions included. The dash-dot-dot line indicates the linear interpolation. For the mechanical mixture model $\kappa$ is slightly upwards protruding.

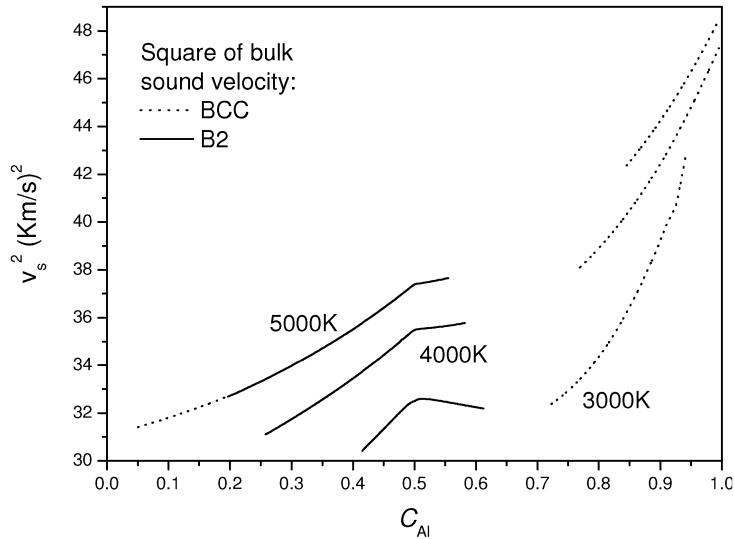

FIG. 9: The square of bulk sound velocity under 30GPa of Ni-Al alloys on bcc lattice with vibrational contributions included. Notice in reality the B2 phase should be equilibrating with liquid or fcc phases and the temperatures are for reference only.

To better understand the behavior of alloys mentioned above, it is helpful to decompose the Gibbs free energy of formation into contributions such as internal energy, vibrational entropy, configurational entropy and volume difference times pressure, respectively. The formation Gibbs free energy is defined relative to that of the mixing model, namely, the mechanical mixture of the ingredients as

$$\triangle G(T,P) = G - [c_{Al}G_{Al}(T,P) + (1-c_{Al})G_{Ni}] = \triangle E + P\triangle V - T\triangle S_{vib} - TS_{cvm}. \quad (12)$$

The magnitudes of each partial Gibbs free energy of formation at 2100K and 30GPa are shown in figure 10. The internal energy is the largest contribution, followed by the volume difference and configurational entropy terms. The vibrational entropy difference is much less and almost negligible. The sharp turns in the curves of internal energy and configurational entropy in the ordered single phase region suggest a connection to the "W-shape" of the EOS properties.

Variations of the EOS quantities as functions of temperature, pressure and concentration are determined completely by the Gibbs free energy as a functional of volume $V$ and correlation functions $\xi_i$ (after integrating out other degrees of freedom, say, the concentration $c$). By defining a vector variable $\eta_0 = V$ and $\eta_i = \xi_i, (i = 1, 2, \ldots)$, and using the variational

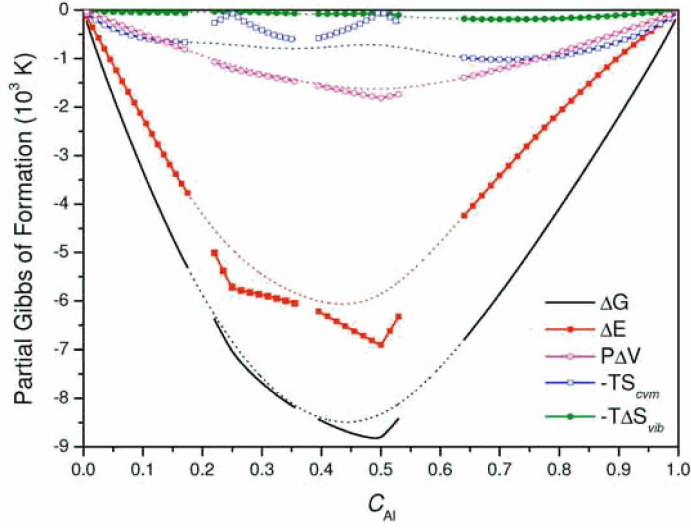

FIG. 10: The partial Gibbs free energy of formation at 30GPa and 2100K for the fcc Ni-Al system. Dotted lines indicate metastable/coexisting phase regions.

condition $\frac{\delta G}{\delta \eta} = 0$, one obtains

$$\left(\frac{\partial \eta}{\partial T}\right)_i = \sum_j \left(\frac{\partial^2 G}{\partial \eta \partial \eta}\right)_{i,j}^{-1} \left[\frac{\partial H}{T \partial \eta} - \frac{\partial^2 H}{\partial T \partial \eta} + T \frac{\partial^2 S}{\partial T \partial \eta}\right]_j, (i, j = 0, 1, 2, \ldots), \quad (13)$$

where $H$ is enthalpy and $S$ the entropy including vibrational contributions. $\left(\frac{\partial^2 G}{\partial \eta \partial \eta}\right)^{-1}$ is the inverse of the Hessian matrix with subscripts $i$ and $j$ labeling matrix elements. Using this relation, the heat capacity $C_P$ is given by

$$C_P = \left(\frac{dH}{dT}\right)_{c,P} = \frac{\partial H}{\partial T} + \sum_i \left(\frac{\partial H}{\partial \eta_i}\right)_{c,P} \left(\frac{\partial \eta_i}{\partial T}\right)_{c,P}$$
$$= \frac{\partial H}{\partial T} + \sum_{ij} \left(\frac{\partial H}{\partial \eta}\right)_i \left(\frac{\partial^2 G}{\partial \eta \partial \eta}\right)_{i,j}^{-1} \left[\frac{\partial H}{T \partial \eta} - \frac{\partial^2 H}{\partial T \partial \eta} + T \frac{\partial^2 S}{\partial T \partial \eta}\right]_j. \quad (14)$$

Here the subscripts $c, P$ indicate that both composition $c$ and pressure $P$ are constants. Similarly, the thermal expansion coefficient $\alpha$ is expressed as

$$\alpha = \frac{1}{V}\left(\frac{\partial V}{\partial T}\right)_P = \frac{1}{V}\sum_i \left(\frac{\partial^2 G}{\partial \eta \partial \eta}\right)_{0,i}^{-1} \left[\frac{\partial H}{T \partial \eta} - \frac{\partial^2 H}{\partial T \partial \eta} + T \frac{\partial^2 S}{\partial T \partial \eta}\right]_i, \quad (15)$$

and Grüneisen parameter $\gamma$ is related to the correlation functions via Eq.(11) where the heat capacity at constant volume $C_V$ is given by

$$C_V = \left(\frac{dE}{dT}\right)_{c,V} = \frac{\partial E}{\partial T} + \sum_{ij} \left(\frac{\partial E}{\partial \xi}\right)_i \left(\frac{\partial^2 F}{\partial \xi \partial \xi}\right)_{i,j}^{-1} \left[\frac{\partial E}{T \partial \xi} - \frac{\partial^2 E}{\partial T \partial \xi} + T \frac{\partial^2 S}{\partial T \partial \xi}\right]_j, \quad (16)$$

where $F$ is the Helmholtz free energy and $E$ the internal energy. These relations indicate that the "W-shape" is directly related to the behavior of inverse Hessian matrix. Some of its elements dominate the detailed thermodynamical behavior of alloys, mainly from the variation of Gibbs free energy with respect to correlation functions. In contrast to the previously mentioned properties, the compressibility is determined only by the variation of free energy with respect to volume,

$$\kappa^{-1} = V\frac{\partial^2 F}{\partial V^2}, \tag{17}$$

and it is unrelated to the correlation functions, and the "W-shape" is not found for $\kappa$, as shown in figure 8, so as for the bulk sound velocity on bcc-based phases (cf. FIG.9). As the "W-shape" composition dependence is governed mainly by the general behavior of the inverse of Hessian matrix with respect to correlation functions for order-disorder systems, the conclusions drawn here should be valid also for other systems, e.g., interstitial alloys and mineral crystals.

## IV. CONCLUSION

The variation of EOS quantities as functions of concentration and temperature as calculated with *ab initio* ECIs was presented. The "W-shape" in the composition dependence around stoichiometry is observed for several important properties. Analysis shows that this kind of behavior is related to the behavior of the inverse of Hessian matrix with respect to correlation functions. This explains the similarity in behavior of the heat capacity and the thermal expansion coefficient, and the absence of the "W-shape" near stoichiometry for the compressibility and the bulk sound velocity. The strong composition dependence near stoichiometry due to configurational corrections has not received much attention before and may be helpful for understanding subtle phenomena in alloys and mineral crystals. The configurational corrected Grüneisen parameter is also shown to have strong composition dependence near stoichiometry around $T_c$. This suggests that the EOS of order-disorder systems is much more complicated than previously expected and that configurational effects cannot be neglected. In addition, the variation of the bulk sound velocity in figure 9 is much attractive. Similar phenomena can be expected for transverse or longitudinal vibration modes, and might be useful for deep understanding of the propagation of seismic waves in mineral crystals.

**Acknowledgments**

This work was supported by the National Advanced Materials Committee of China. And the authors gratefully acknowledge the financial support from 973 Project in China under Grant No. G2000067101. Part of this work was performed under the inter-university cooperative research program of the Laboratory for Advanced Materials, Institute for Materials Research, Tohoku University.

---